\documentclass[english,preprint, aps]{revtex4-1}
\usepackage[LGR,T1]{fontenc}
\usepackage[latin9]{inputenc}
\usepackage{color}
\usepackage{array}
\usepackage{textcomp}
\usepackage{multirow}
\usepackage{amsmath}
\usepackage{amssymb}
\usepackage{graphicx}

\makeatletter

\DeclareRobustCommand{\greektext}{%
  \fontencoding{LGR}\selectfont\def\encodingdefault{LGR}}
\DeclareRobustCommand{\textgreek}[1]{\leavevmode{\greektext #1}}
\ProvideTextCommand{\~}{LGR}[1]{\char126#1}

\providecommand{\tabularnewline}{\\}

\makeatother

\usepackage{babel}
\begin{document}
\title{The Observation of Jet Azimuthal Angular Decorrelations at FCC-ep}
\author{\.{I}. Ho\c{s}}
\email{ilknur.hos@iuc.edu.tr, ilknur.hos@cern.ch}

\affiliation{\noindent Istanbul University-Cerrahpasa, Faculty of Engineering,Engineering
Sciences, 34320, Istanbul, Turkey}
\altaffiliation{and also CERN-CMS Experiment, Esplanade des Particules 1, P.O. Box 1211 Geneva 23, Switzerland }

\author{S. Kuday}
\email{corresponding author: Skuday@ankara.edu.tr, kuday@cern.ch}

\affiliation{Ankara University, Faculty of Science, Department of Physics, 06100,
Ankara, Turkey}
\altaffiliation{and also CERN-ATLAS Experiment, Esplanade des Particules 1, P.O. Box 1211 Geneva 23, Switzerland }

\author{H. Sayg\i n}
\email{hasansaygin@aydin.edu.tr}

\affiliation{Istanbul Ayd\i n University, Application and Research Center For Advanced
Studies, 34295, Istanbul, Turkey}
\keywords{QCD, Jets, MN-Jets, BFKL, Small-x}
\begin{abstract}
Higher collision energies at future colliders may eventually lead
to the falsification of standard fixed-order perturbation theory and
linear evolutions due to non-linear structure of QCD at small-x. New
physics researches \textcolor{black}{that are strictly based on accurate
jet measurements will undoubtedly have this observation known as BFKL
effect via angular jet decorrelations taking into account the Mueller-Navelet
jets. As one of the frontier colliders, FCC-ep, has a great observation
potential on parton densities through asymmetrical collisions. We
aim to test the observability of azimuthal angular jet decorrelations
with the recent event generators (HERWIG, PYTHIA) at the generator
and detector level for FCC-ep centre of mass energies \textsurd s
= 3.5 TeV in proton-electron collisions. Jets are reconstructed by
the anti-$k_{T}$ algorithm ($R=0.5$), with $p_{T}>15$GeV and selected
in the range of $|\eta|<6$. Relevant pseudo-rapidity regions have
been analyzed with the azimuthal-angle difference between Mueller-Navelet
Jets ($\Delta\text{\textgreek{F}}$) in the pseudo-rapidity separation
($\Delta\eta$) and the distributions of $<cos(n(\pi-\Delta\phi))>$
are presented in comparison as the res}ult.
\end{abstract}
\maketitle

\section{INTRODUCTION}

Following the collision of protons in a collider, high-energy quark
and gluon production emerges jets as the sprays of quarks. The process
that forms the jet structures based on the quark fragmentation is
not fully understood in perspective of perturbation theory. \textcolor{black}{The
strong interaction between quarks and gluons (partons) is defined}
by the theory of Quantum Chromodynamics (QCD). It is known that jet
formation collects important contributions from QCD effects as a complex
multiscale problem. The behavior of the QCD interactions with respect
to different momentum scales can be considered as one of the most
puzzling questions \textcolor{black}{within the Standard Model (SM)
theory. If one is considering an experimental setup to collide energetic
non-point particles (e.g: protons), Parton Distribution Functions
(PDFs) should be considered to calculate physical observables. In
experimental data analysis, Monte Carlo (MC) event generators use
built-in PDFs that are strictly depended on momentum scales and can
be recalculated for each center of mass energies to reconstruct background
data. However, recent observations show that as the collision energies
and collected data increase, the kinematic observables reveal some
anomalies in comparison of background and signal \citep{atlas-new-1,atlas-new}.
Especially in large pseudo-rapidity distributions, ATLAS and CMS Collaborations
announced that a good agreement between theory and experiment can
be provided only if the multiple MC generators are assigned for the
similar analysis \citep{CMS-kFactor,CMS-AzDecPaper}. In the past
experiments (e.g: D0), a similar effect was hinted with \textsurd s
= 1.8 TeV, 1800 and 630 GeV at Fermilab Tevatron \citep{D0-1,D0-2}.
In those studies, pseudo-rapidity interval $(\text{\textgreek{D}\textgreek{h}}$)
-as an important variable- is selected up to 6, to limit the observation
of the decorrelation effect. Another recent study performed at \textsurd s
= 100 TeV pp collisions planned for Future Circular Collider (FCC)
\citep{FCC-2} and Circular Electron Positron Collider (CEPC) \citep{CEPC},
shows the BFKL effect becomes significant at the higher-collision
energies \citep{Acta}.}

\textcolor{black}{In this work, we aim to present the observation
of azimuthal angular jet decorrelations through multiple MC generators,
namely HERWIG 7 \citep{herwig} and PYTHIA 8 \citep{pythia8}, at
FCC-ep \citep{FCC-ep}. PYTHIA and HERWIG are both multifunctional
event generators been used for the simulation of high-energy particle
physics events. They are most common event generators used in recent
analyses to define hadronization process and simulation of parton
distributions, initial and final state parton showers, particle decays,
\dots{} etc. PYTHIA allows one to select different incoming beam particles
and has $p_{T}$ ordered parton showers. While different PDF set selections
are}\textcolor{blue}{{} }\textcolor{black}{possible to be used, the
default Parton distribution is Coordinated Experimental Project on
QCD (CTEQ) 5L. In this paper, NNDPDF 2.3 and NNPDF 4.0 pdf sets are
used for both generators with multiple runs \citep{NNpdf23,NNpdf40}}\textcolor{blue}{.}\textcolor{black}{{}
HERWIG also allows the selection of lepton-lepton, lepton-hadron and
hadron-hadron collisions. It uses angular-ordered parton showers in
QCD jet evolution and cluster hadronization model for the hadronization
process. }

As a part of the huge project in the FCC framework, ep collider offers
asymmetrical collisions to researchers to analyze topics such as high
precision QCD, Top\&Electroweak Physics, Supersymmetry and Higgs Physics.
According to recent concept design report of FCC \citep{FCC-CDR},
it is planned to be built on 80-100 km tunnel under CERN campus to
reach the 50 TeV proton beam energy. For the electron beam, it is
aimed to reach 60 GeV energy with boosting particles in the energy
recovery linac. In the physics program of FCC-ep collider that became
evident before CDR report \citep{FCC-1,FCC-2}, QCD studies take an
utmost important place as the complementary to hadron collider studies. 

The outline of the paper is as follows: a numerical approach to the
problem using \textcolor{black}{the Balitsky-Fadin-Kuraev-Lipatov
(BFKL) \citep{BFKL,BFKL-2,BFKL-3}} and\textcolor{black}{{} Dokshitzer,
Gribov, Lipatov, Altarelli, Parisi (DGLAP) \citep{DGLAP-1,DGLAP-2}}
evolutions is mentioned in the section II with the theoretical considerations.
We explain event generation setup and jet selection to obtain Mueller-Navelet
Jets (MN-Jets) in the section III.\textcolor{blue}{{} }\textcolor{black}{Then
we present the analysis results with our the comments in the section
IV. And in the final section, we basically present the quantitative
outputs of the analysis that} may be possible to observe in the FCC-ep
experiments.

\section{THEORETICAL  CONSIDERATIONS}

In the hadronic c\textcolor{black}{ollisions, Mueller-Navelet Jets
(MN-Jets) are some jets that carry the longitudinal momentum fraction
of their parent hadrons in the forward direction and that cause a
large pseudo-rapidity separation between each others. If $k_{1}$
and $k_{2}$, the transverse momenta of the forward jets, can be measured,
the total collision energy would be sufficiently large to observe
MN-Jets in the large pseudo-rapidity interval $\triangle\eta\sim ln(\hat{s}/k_{1}k_{2})$
where }\textit{\textcolor{black}{$\hat{s}$}}\textcolor{black}{{} the
partonic center of mass energy. On the other hand, one can explain
jets in the fixed-order perturbative QCD calculations considering
a fixed value for the running strong coupling $\alpha_{s}$. More
specifically, in order to calculate cross section, one should obtain
the partonic momenta from structure functions within the energy scale
$Q^{2}$ and solve the DGLAP equation as mentioned in ref \citep{DGLAP-1,DGLAP-2}.
Recently, these structure functions have been well-studied within
the PDF studies solving the DGLAP equation that allows resummation
of the large logarithms coming from the strong ordering between the
hadrons scale and the jets transverse momenta using mathematical methods
\citep{DGLAP-3}. Note that recent MC event generators use built-in
PDF datasets although various calculation tools are developed based
on the DGLAP analytical solutions. However, in the DGLAP perspective,
a dijet is correlated change of parton densities with varying spatial
resolution of the detector. With contrary the observations, that lead
to end up with low $p_{T}$ emissions via strong ordering of transverse
momenta.}

\textcolor{black}{In the high-energy regime, the BFKL approach states
that a dijet can be decorrelated with large parton emissions and allows
the resummation of terms with $\alpha_{s}log(1/x)^{n}$at leading
(LL) and next-leading (NLL) logarithmic accuracies. Thus, one can
calculate the cross section values that are independent of the parton
densities. However for higher accuracy one should calculate the NLL-BFKL
\citep{Ducloue-1,Ducloue-2} predictions since it is reported LL-BFKL
is underestimating data \citep{D0-7}. Unfortunately, NLL calculations
are beyond our scope of this work due to technical limitations.}

\textcolor{black}{One can calculate the normalized MN-Jets cross section
analytically as a function of azimuthal-angle difference ($\Delta\phi$)
with $p_{T}>p_{(T,min)}$ in Fourier series expansion as follows \citep{Duca,Stirling,murzin}}\textcolor{blue}{:}

\begin{equation}
\frac{1}{\sigma}\frac{d\sigma}{d(\Delta\phi)}(\Delta y,\:p_{(T,min)})=\frac{1}{2\pi}[1+2\stackrel{}{\underset{n=1}{\overset{\infty}{\sum}}C_{n}(\Delta y,\:p_{(T,min)})\,cos(n(\pi-\Delta\phi))}]
\end{equation}

Here, $C_{n}$ parameters are Fourier coefficients and equal to average
cosines of the deco\textcolor{black}{rrelation angle, $<cos(n(\pi-\Delta\phi))>$,
where $\Delta\phi=\phi_{1}-\phi_{2}$ is the difference between azimuthal
angles of MN-Jets. Phenomenologically the reason of chosing average
cosine obse}rvable has a direct effect on differential MN-Jets cross
section as well as it's a kinematically measurable variable. 

\section{EVENT AND JET SELECTION}

PYTHIA 8 (version of 8.243) and HERWIG 7 (version of 7.1.2) are used
to generate event\textcolor{black}{s with electron-proton collisions
at \textsurd s = 3.5 TeV. For HERWIG event production, we set the
center of mass energy to 3464.1 GeV with 50 TeV proton beam and 60
GeV electron beam. In the final state, lepton + jet is selected allowing
on shell production for all stable particles. We also implemented
QCD 2->2 and Deep Inelastic Scaterring (DIS) processes. We observed
that although multiparticle interactions are allowed and standard
coupling orders are considered, those have no significant effects
on our analysis. For events hadronized with PYTHIA 8, MADGRAPH \citep{madgraph}
(version of 3.4.1) has been used to collide the electron and proton
beams with the center of mass energies 60 GeV and 50 TeV, respectively.
In the final state, lepton + boson + X process has been selected.
Then events generated by both MC generators are used to reconstruct
jets with anti-$k_{T}$\citep{anti-kT} jet algorithm (cone radius
of 0.5) within FASTJET (version of 3.3.0) \citep{fastjet}. As the
detector implementation, we utilized DELPHES v.3-4-0}\textcolor{blue}{{}
}\textcolor{black}{\citep{Delphes} that has the recent FCC-hh detector
specifications in parallel to experimental aims. Thus, we set the
radius of the magnetic field coverage to 1.5 m in 4 T magnetic field.
We have used standard efficiency formulas/algorithms in electron,
muon and hadron tracking}\textcolor{blue}{{} }\textcolor{black}{\citep{NewDelphes}.
As the preselection criteria that is considered highly inclusive,
events with at least two jets are used and jets are required to pass
$p_{T}$ cut of 10 GeV and to be in the pseudo-rapidity region of
$|\eta|<7.$}

\textcolor{black}{In the analysis, the following criteria and steps
are required to select the MN-Jets:}
\begin{itemize}
\item \textcolor{black}{$p_{T}$ higher than 15 GeV }
\item \textcolor{black}{in the pseudo-rapidity region of $|\eta|<6.$ }
\item \textcolor{black}{apply pseudo-rapidity ordering of jets for each
event }
\item \textcolor{black}{choose the jets with highest pseudo-rapidity and
lowest pseudo-rapidity value }
\item \textcolor{black}{name them the most forward jet and the most backward
jet (MN-Jets), respectively. }
\end{itemize}
\textcolor{black}{Two jets in each event with the largest pseudo-rapidity
separation are obtained. Some kinematic distributions of these jets
are produced to see the selection of jets at detector level. In Figure
1 top plot, transverse momentum ($p_{T}$) distributions of forward
and backward jets are plotted with respect to each other. $p_{T}$
higher than 15 GeV is applied to these jets as mentioned above. There
is no enrty for the backward jets in the region of $\eta_{2}>1$ due
to the assymetry of collision. In Figure 1, bottom left plot shows
pseudo-rapidity ($\eta$) distributions of MN-Jets telling the most
of the jets are back-to-back in the plane of $\eta$, while right
plot gives the phi ($\Phi$) distribution of these jets indicating
most of the jets are in back-to-back in $\Phi$ plane. }

\textcolor{black}{The number of events and number of jets before and
after cuts are presented in Table 1 for PY}THIA 8 and HERWIG 7 separately.
We believe as a reason of that the different jet treatments from both
generators are due to optimization of FASTJET algorithm that is used
via DELPHES settings. 

\begin{table}
\caption{Number of events and number of jets before and after cuts at \textsurd s
= 3.5 TeV}

\begin{centering}
\begin{tabular}{|c|c|c|c|c|}
\hline 
@\textsurd s = 3.5 TeV &  & Before Cuts & After Cuts{*} & After MN-Jets Selection Criteria\tabularnewline
\hline 
\hline 
PYTHIA 8 &  & \multicolumn{3}{c|}{}\tabularnewline
\hline 
\multirow{1}{*}{Number of Events} &  & 6.24692e+06 & 3.86211e+06 & 3.86208e+06\tabularnewline
\hline 
\multirow{2}{*}{Number of Jets} & Gen. Level & 3.13332e+07 & 2.50009e+07 & 8.25519e+06\tabularnewline
\cline{2-5} \cline{3-5} \cline{4-5} \cline{5-5} 
 & Det. Level & 1.18196e+07 & 9.97976e+06 & 9.97967e+06\tabularnewline
\hline 
\hline 
HERWIG 7 &  & \multicolumn{3}{c|}{}\tabularnewline
\hline 
\multirow{1}{*}{Number of Events} &  & 3.845e+07 & 1.8e+07 & 1.758e+06\tabularnewline
\hline 
\multirow{2}{*}{Number of Jets} & Gen. Level & 7.38e+08 & 473785 & 33817\tabularnewline
\cline{2-5} \cline{3-5} \cline{4-5} \cline{5-5} 
 & Det. Level & 1.817e+08 & 4.3767e+07 & 1.38697e+07\tabularnewline
\hline 
\end{tabular}
\par\end{centering}
{\footnotesize{}\smallskip{}
{*} Events with at least 2 jets \& jet pT > 15 GeV}{\footnotesize\par}
\end{table}

\begin{table}
\caption{Number of jets at each $\Delta\text{\textgreek{F}}$ distribution
at \textsurd s = 3.5 TeV}

\centering{}%
\begin{tabular}{|c|c|c|c|c|}
\hline 
 &  & $|\text{\textgreek{D}}\eta|<3.$ & $3.<|\text{\textgreek{D}}\eta|<6.$ & $6.<|\text{\textgreek{D}}\eta|<9.$ \tabularnewline
\hline 
\hline 
\multirow{2}{*}{PYTHIA 8} & Gen. Level & 4518190 & 2240349 & 1496647\tabularnewline
\cline{2-5} \cline{3-5} \cline{4-5} \cline{5-5} 
 & Det. Level & 9556768 & 341130 & 81769\tabularnewline
\hline 
\multirow{2}{*}{HERWIG 7} & Gen. Level & 33886 & 580 & 44\tabularnewline
\cline{2-5} \cline{3-5} \cline{4-5} \cline{5-5} 
 & Det. Level & 67892 & 283 & 22\tabularnewline
\hline 
\end{tabular}
\end{table}

\begin{figure}
\includegraphics[scale=0.35]{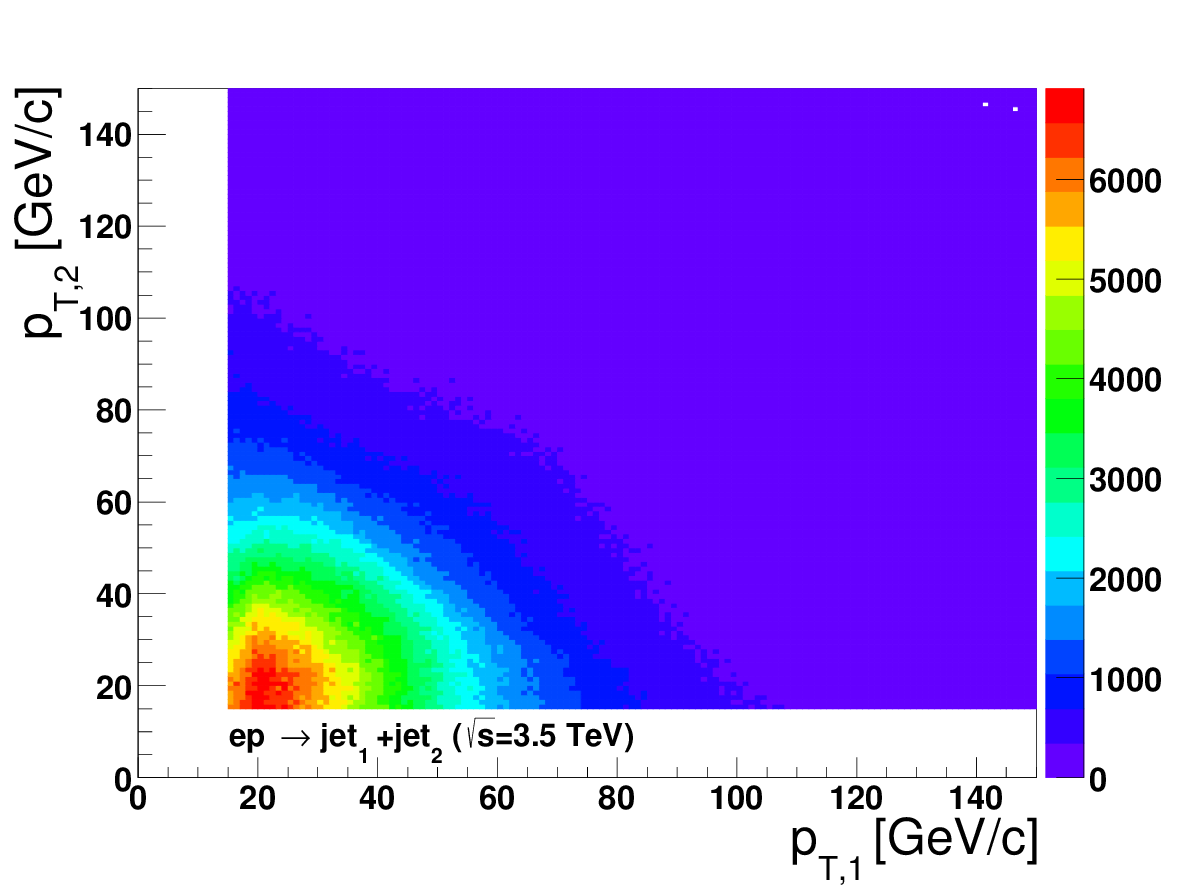}\,\quad{}\includegraphics[scale=0.35]{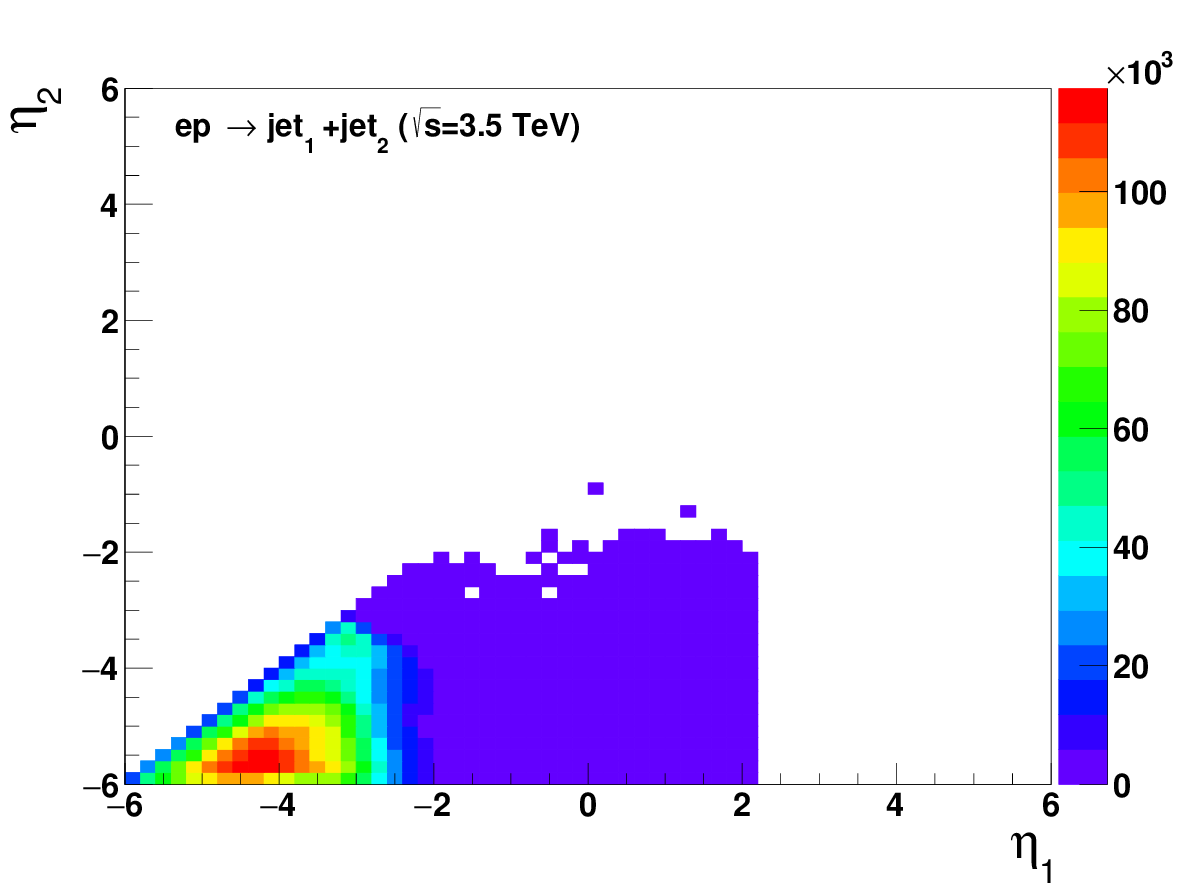}\includegraphics[scale=0.35]{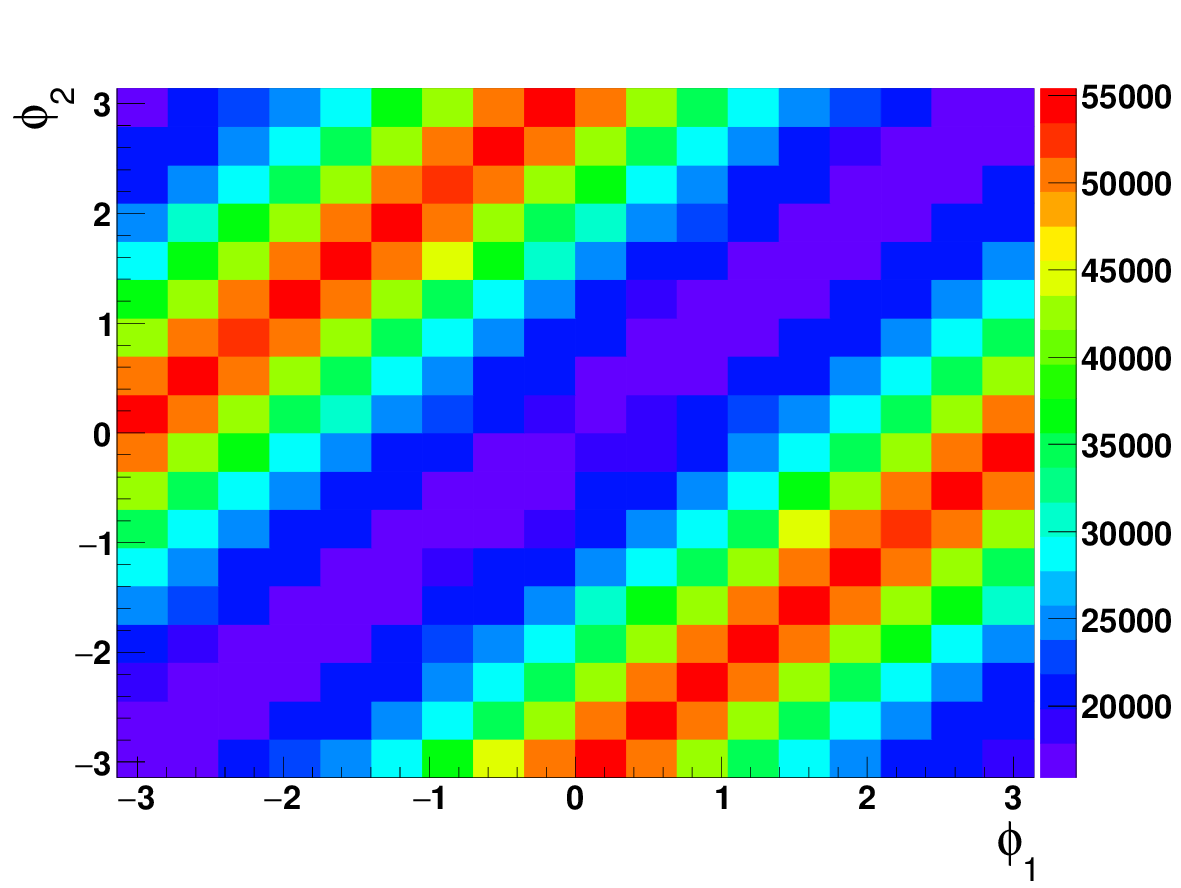}

\caption{$p_{T}$ , $\eta$ and $\Phi$ distributions of forward jet vs backward
jet for PYTHIA 8 at detector level. }
\end{figure}

\section{ANALYSIS}

The azimuthal-angle difference between \textcolor{black}{MN-Jets ($\Delta\text{\textgreek{F}}$)
as} a function of the pseudo-rapidity separation is plotted in Figure
2. Top plots show the distribution for PYTHIA 8 while bottom plots
are repres\textcolor{black}{enting the distribution for HERWIG 7.
The distributions are plotted for three pseudo-rapidity separations:
$|\text{\textgreek{D}}\eta|<3.$, $3.<|\text{\textgreek{D}}\eta|<6.$,
and $6.<|\text{\textgreek{D}}\eta|<9.$ The shape of the distributions
is reversed from the first binning to the last one for both generator
and detector levels. However it is barely visible for HERWIG 7 due
to the lower statistics. Table 2 shows the number of generetad jets
and detector jets at each $\Delta\text{\textgreek{F}}$ distribution
for PYTHIA and HERWIG at \textsurd s = 3.5 TeV. This table gives the
number of jets in the distributions plotted in Figure 2. The number
of jets decreases with large pseudo-rapidity separation for both MC
generators. The peak of $\Delta\text{\textgreek{F}}$ distribution
decreases and the distribution becomes wider comparing to the distributions
with narrower $\Delta\eta$ with increasing pseudo-rapidity between
jets. }

\textcolor{black}{Figure 3 represents the distribution of $<cos(n(\pi-\Delta\phi))>$for
both MC event generators. Distribution of PYTHIA 8 (top plots) shows
fluctuations as a function of $\text{\textgreek{D}}\eta$ and last
binning has very low statistics. The peaks in the interval $\Delta\eta$
= 4-5 are observed to disappear after selection of updated PDF sets
such as NNPDF4.0. One can conclude that the flatness or skewness of
the distribution is due to statistical fluctuations within the PDF
uncertainities. HERWIG 7 distribution, bottom plots in Figure 3, show
a decrease with increasing of $\text{\textgreek{D}}\eta$ which indicates
a better sign of decorrelation. }

\textcolor{black}{The ratio of $<cos2(\pi-\Delta\Phi)>$ to $<cos(\pi-\Delta\Phi)>$
($\frac{C_{2}}{C_{1}}$ , left plot) and $<cos3(\pi-\Delta\Phi)>$
to $<cos2(\pi-\Delta\Phi)>$ ($\frac{C_{3}}{C_{2}}$ , right plot)
as a function of the pseudo-rapidity separation $\Delta\eta$ are
plotted for PYTHIA 8 (top) and HERWIG 7 (bottom) in Figure 4. The
distribution of HERWIG 7 shows a smooth decrease downwards versus
the higher values of $\Delta\eta$ and last binnings of histograms
are suffering from low statistics. Main reasons for the differences
in those plots are basically due to the unsimilar computational approaches
of event generators as it will be discussed in the next section. }

\begin{figure}
\includegraphics[scale=0.4]{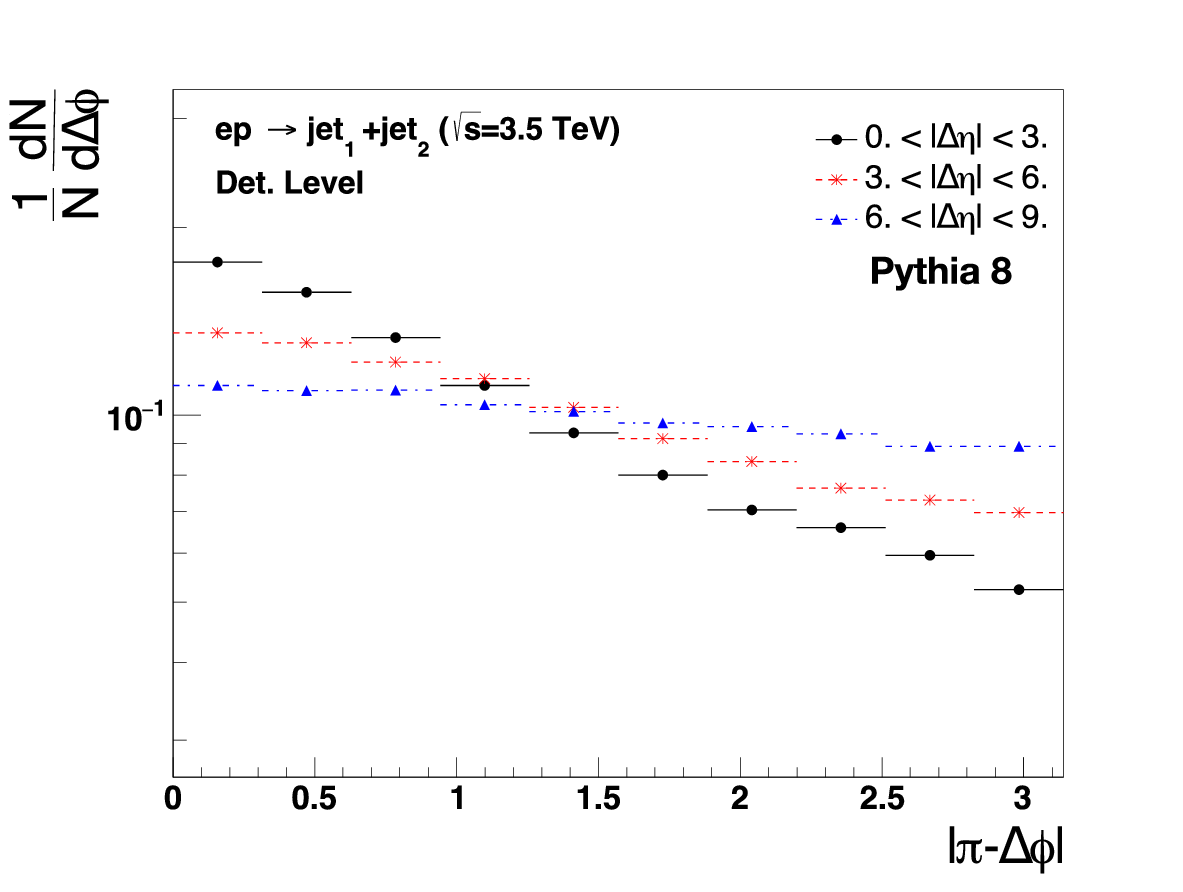}\includegraphics[scale=0.4]{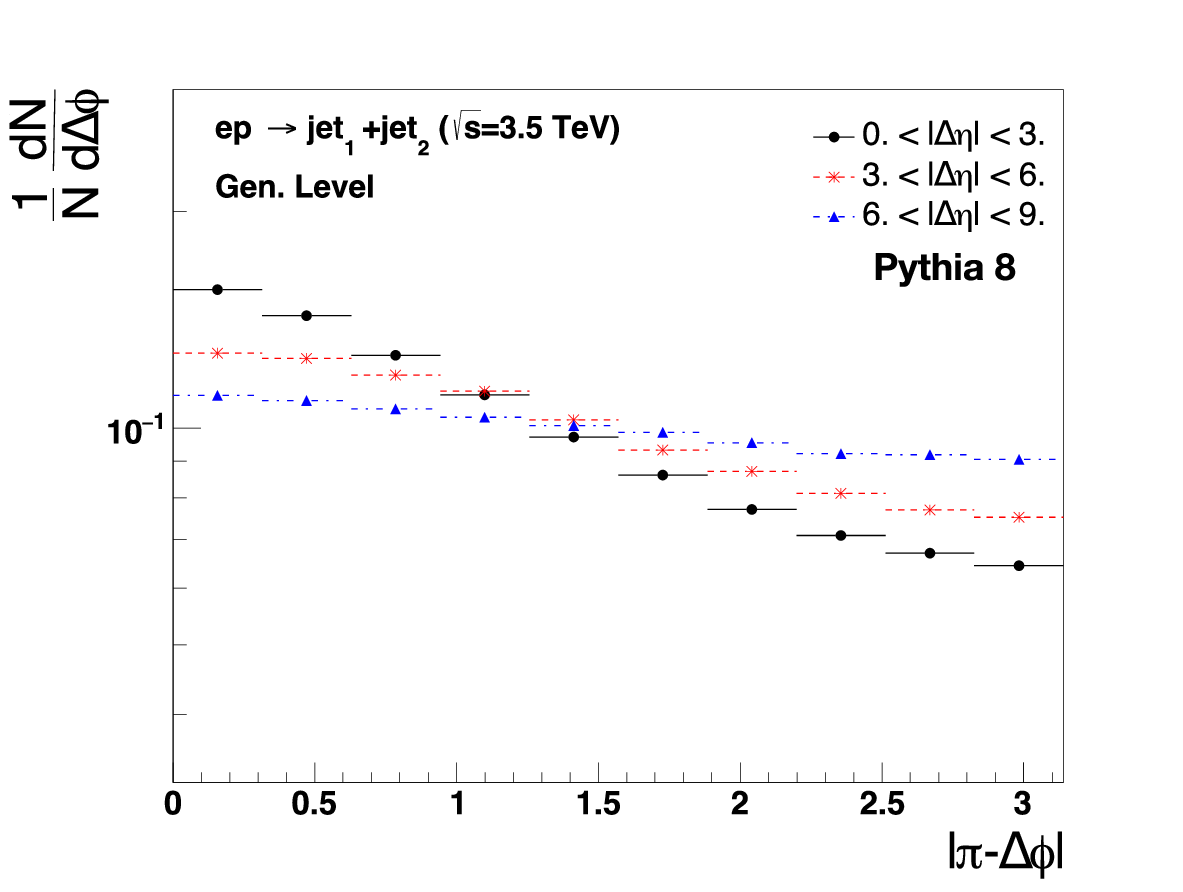}

\textcompwordmark\includegraphics[viewport=0bp 0bp 568bp 427bp,scale=0.4]{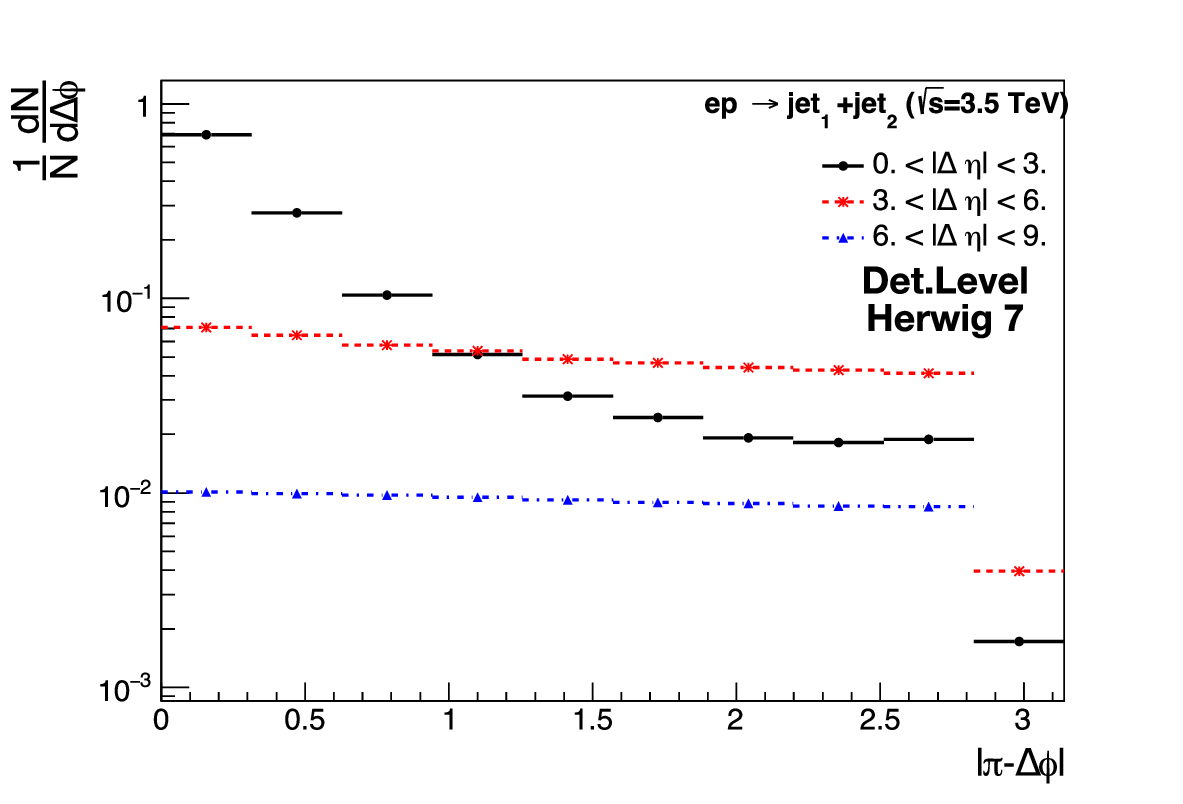}\includegraphics[viewport=0bp 0bp 595bp 427bp,scale=0.41]{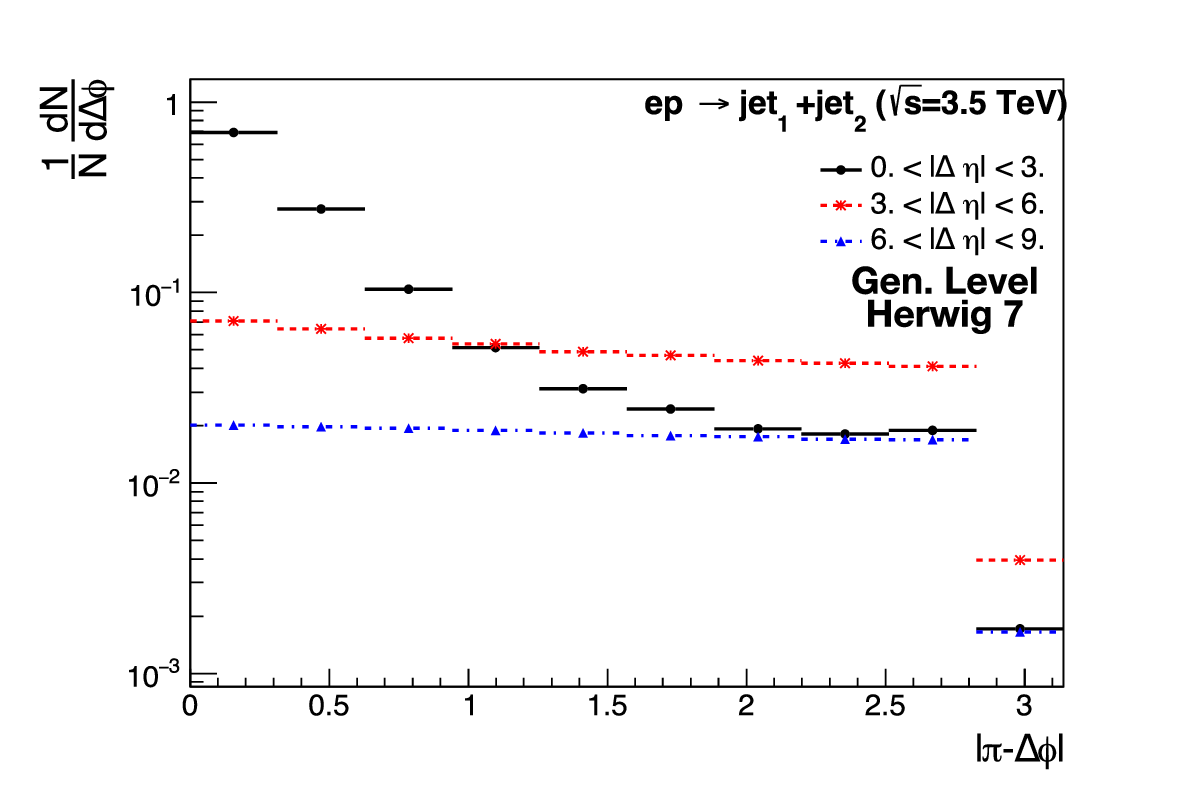}

\caption{The azimuthal-angle difference between \textcolor{black}{MN-Jets ($\Delta\text{\textgreek{F}}$)
in} the pseudo-rapidity of $|\text{\textgreek{D}}\eta|<3.$, $3.<|\text{\textgreek{D}}\eta|<6.$,
and $6.<|\text{\textgreek{D}}\eta|<9.$ at detector level (left) and
generator level (right) for PYTHIA 8 (top) and HERWIG 7 (bottom)}
\end{figure}

\begin{figure}
\includegraphics[viewport=0bp -1bp 568bp 427bp,scale=0.4]{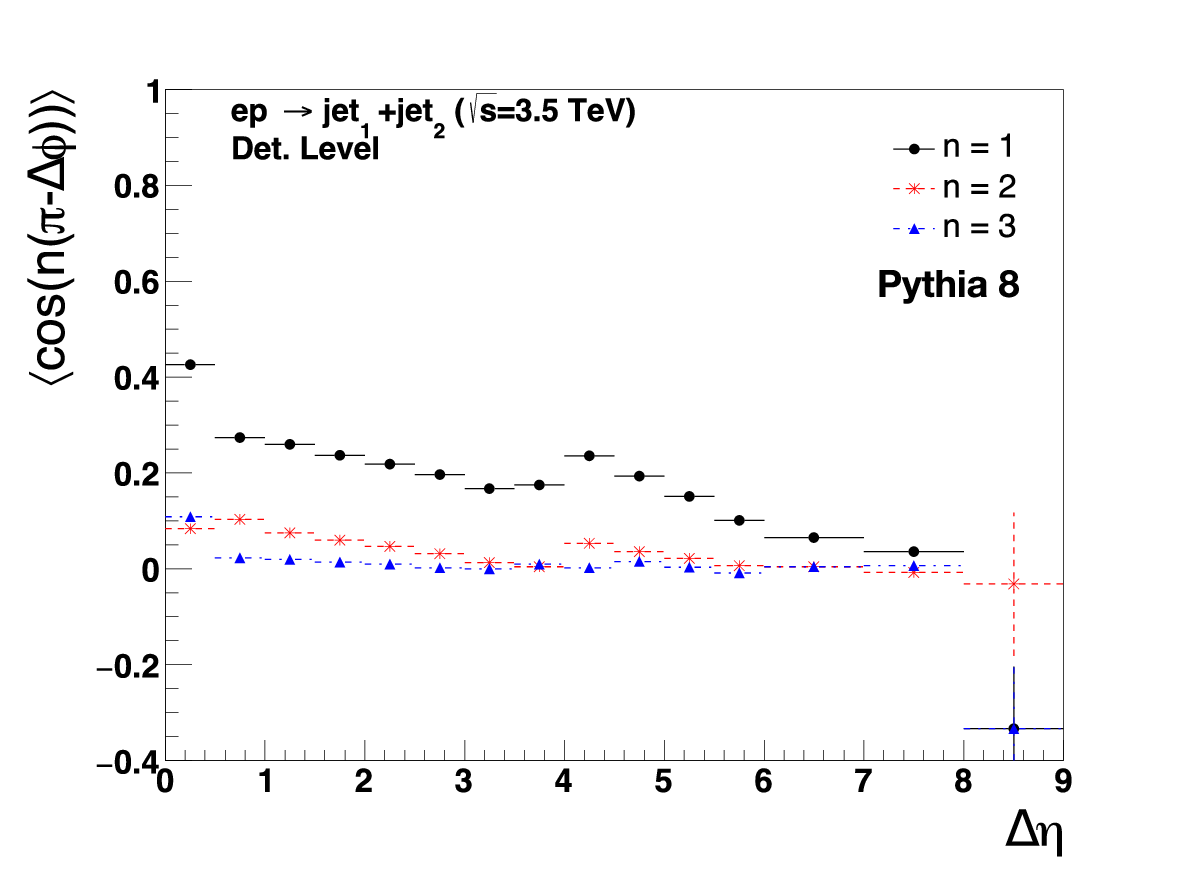}\includegraphics[scale=0.4]{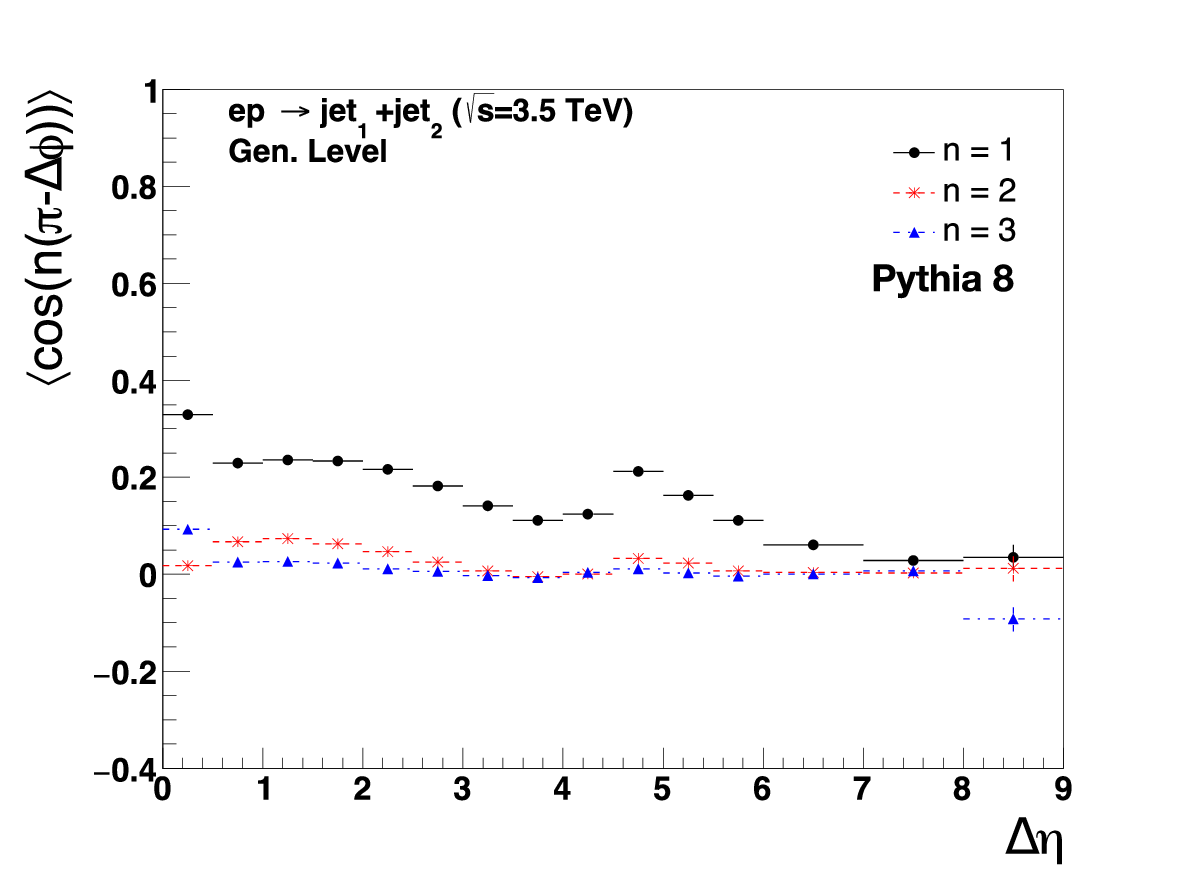}

\textcompwordmark\includegraphics[viewport=0bp 0bp 595bp 427bp,scale=0.4]{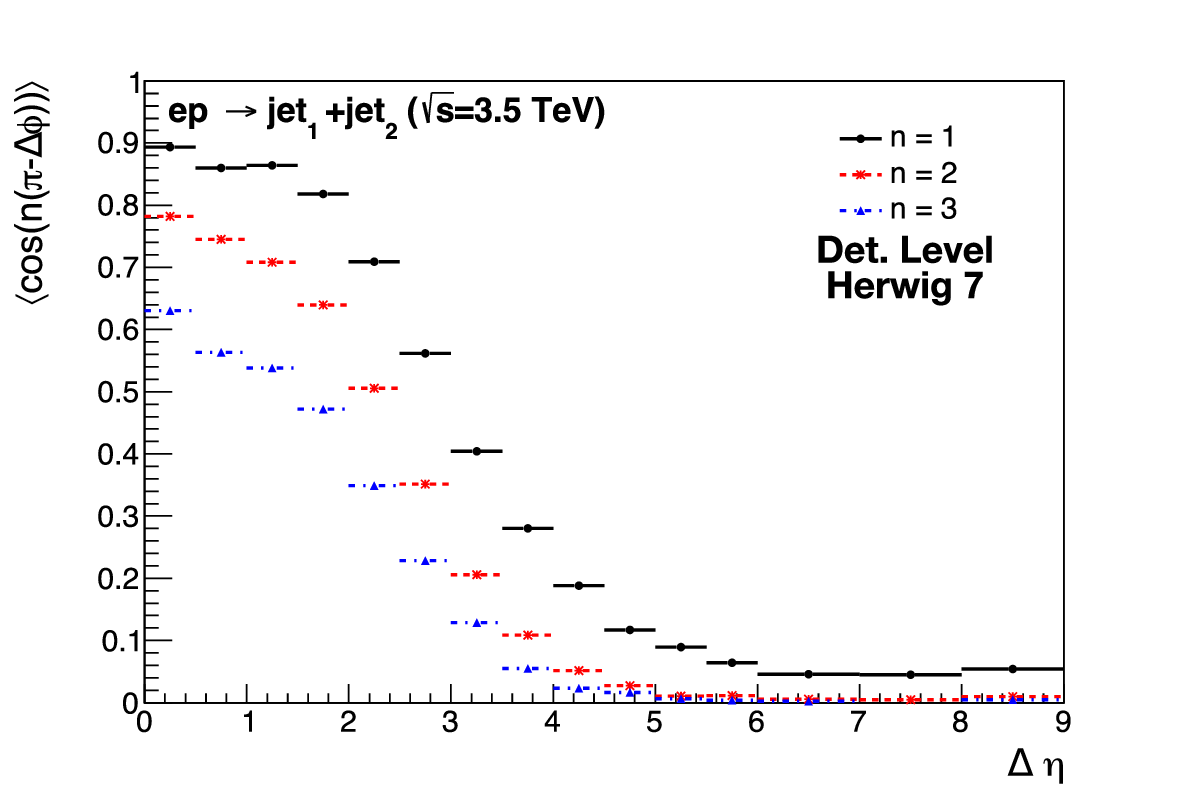}\includegraphics[viewport=10bp 10bp 595bp 427bp,clip,scale=0.34]{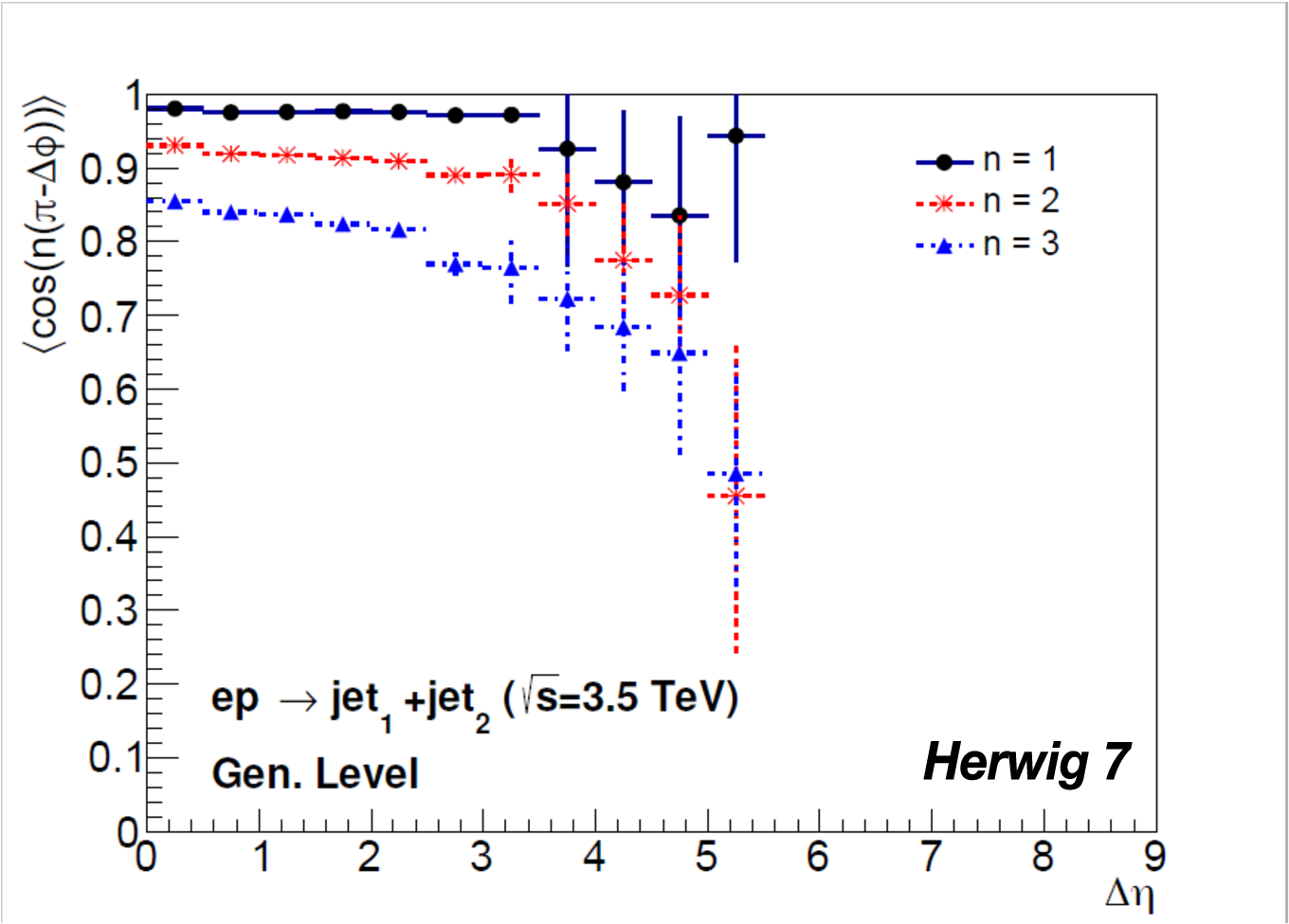}

\caption{$<cos(\pi-\Delta\Phi)>$, $<cos2(\pi-\Delta\Phi)>$ and $<cos3(\pi-\Delta\Phi)>$
as a function of $\Delta\eta$ at detector level (left) and generator
level (right) for PYTHIA 8 (top) and HERWIG 7 (bottom)}
\end{figure}

\begin{figure}
\includegraphics[scale=0.4]{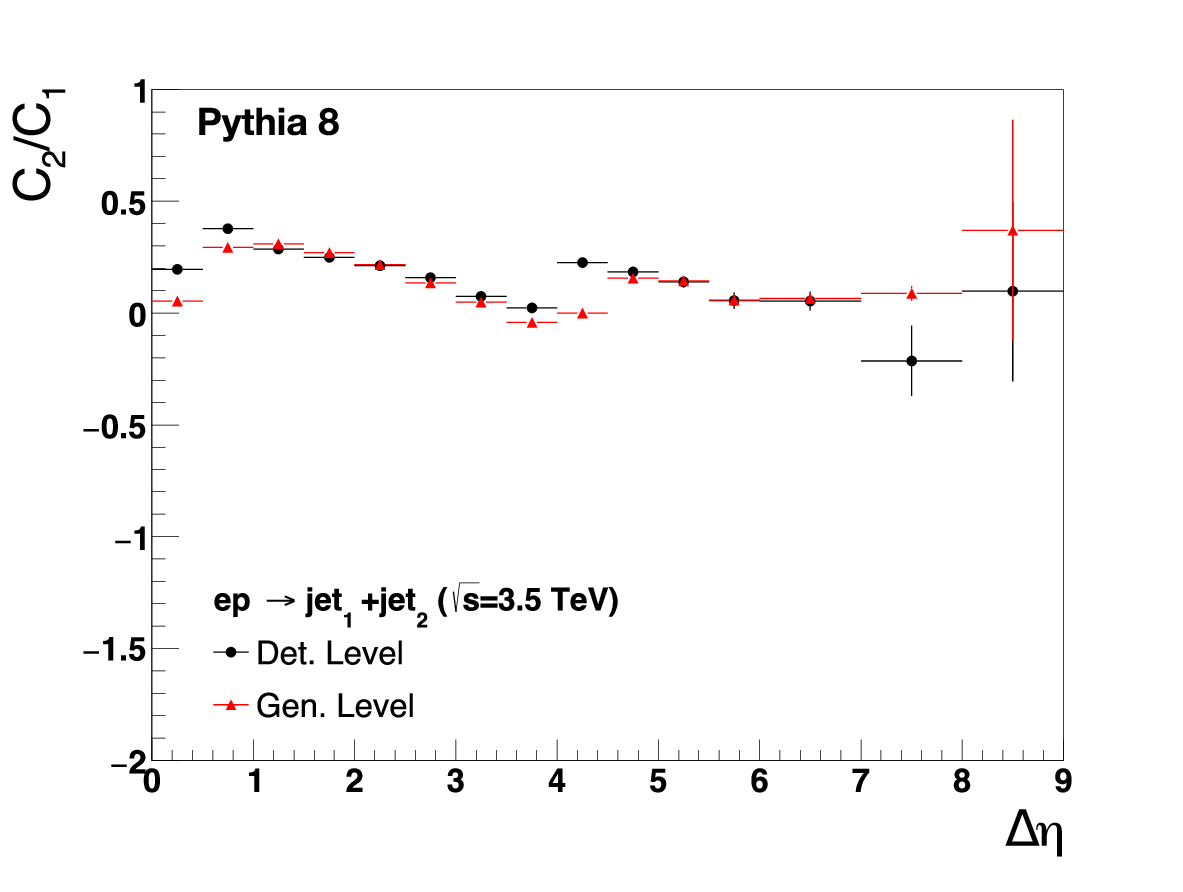}\includegraphics[scale=0.4]{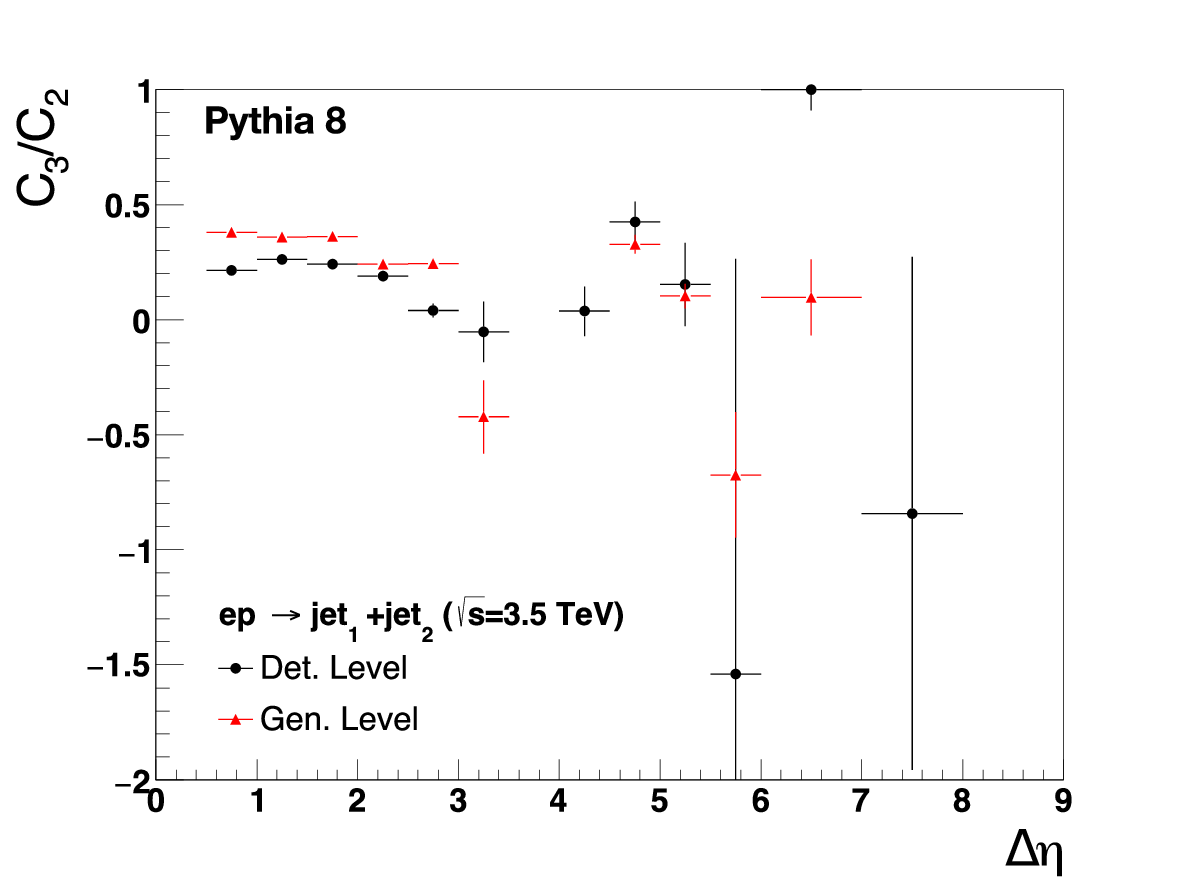}

\includegraphics[scale=0.4]{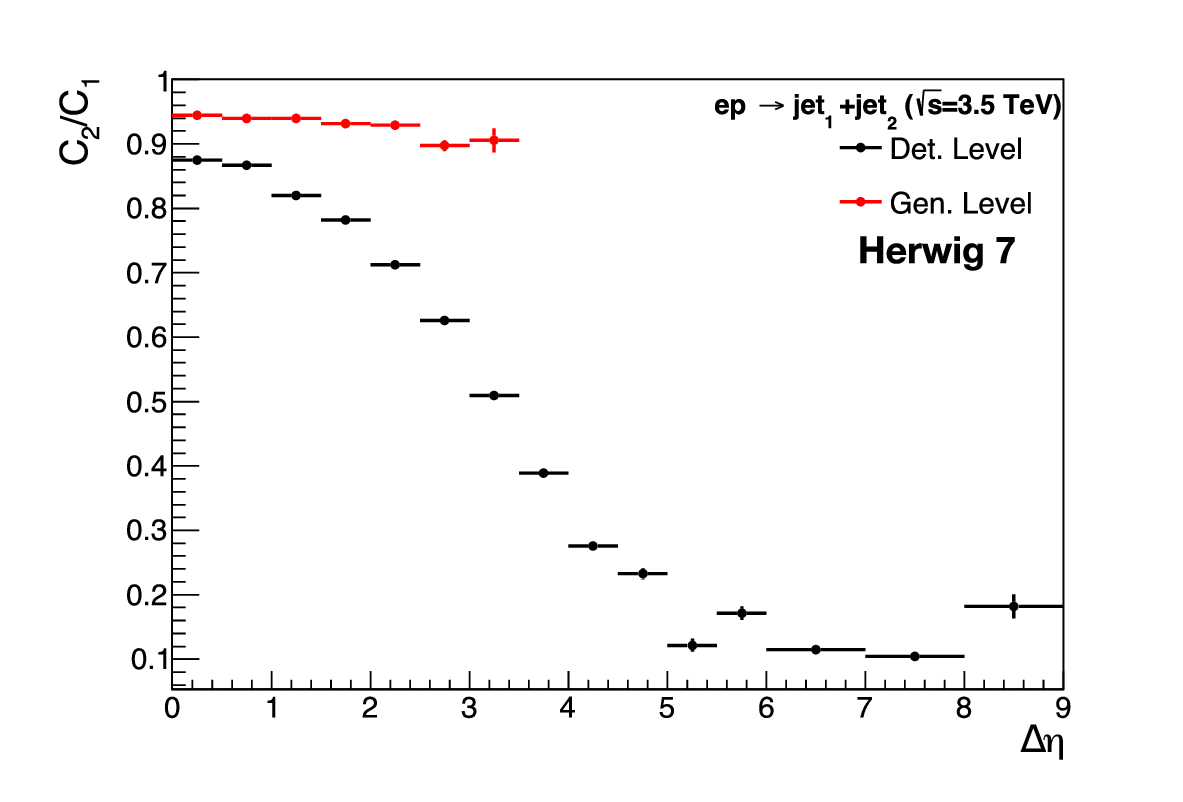}\includegraphics[scale=0.4]{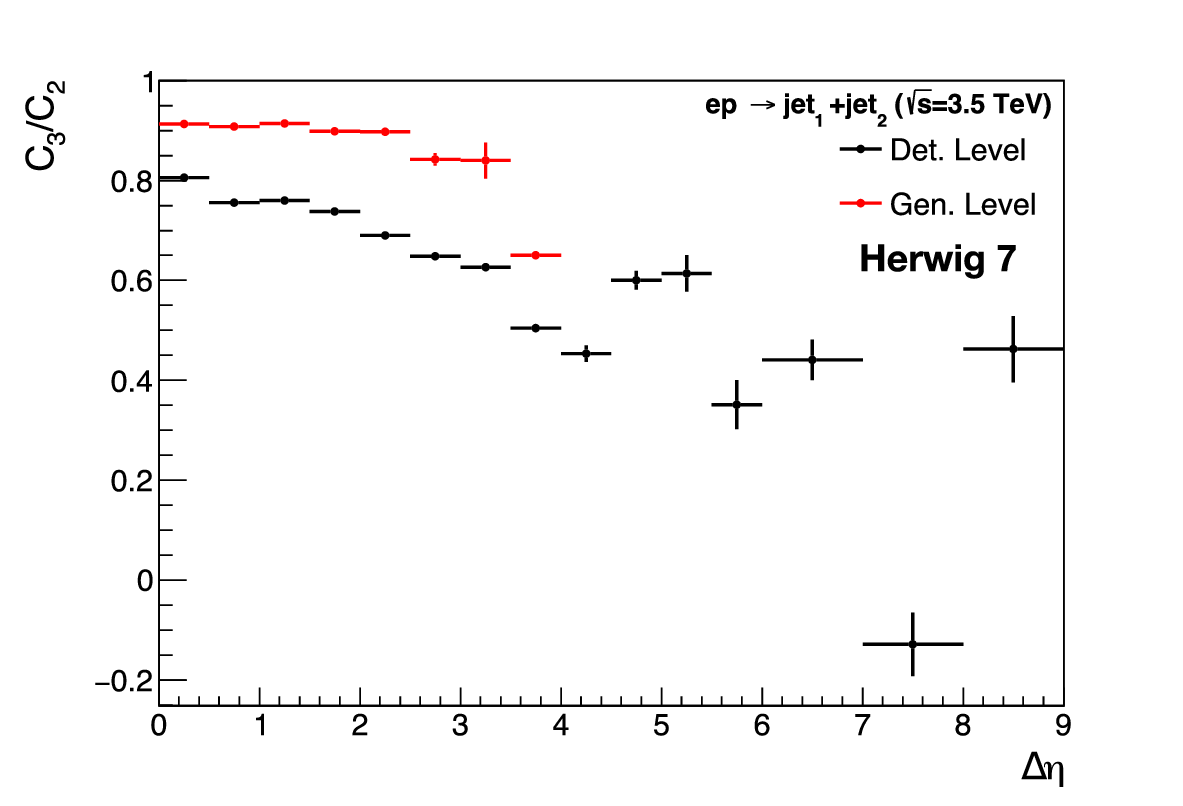}

\caption{Ratio of average cosine $\frac{C_{2}}{C_{1}}$ (left) and $\frac{C_{3}}{C_{2}}$
(right) as a function of $\Delta\eta$ at \textsurd s = 3.5 TeV for
PYTHIA 8 (top) and HERWIG 7 (bottom)}
\end{figure}

\section{CONCLUSION}

We have presented \textcolor{black}{azimuthal angular decorrelation
of most forward and most backward jets in ep collisions at \textsurd s
= 3.5 TeV with PYTHIA 8 and HERWIG 7 MC event generators. Azimuthal-angle
difference between MN-Jets ($\Delta\text{\textgreek{F}}$), average
cosine value of $\Delta\text{\textgreek{F}}$ ($<cos(n(\pi-\Delta\phi))>$)
for $n=1,2,3$ and ratio of $<cos(n(\pi-\Delta\phi))>$ for different
$n$ values as a function of the pseudo-rapidity separation up to
9 are plotted. It's produced around $3\times10^{7}$events for both
MC generators that roughly corresponds to the integrated luminosity
at the order of $(nb)^{-1}$. We have not observed a significant change
in the behaviour of event generators for different multi-parton interaction
(MPI) schemes. At this point, we should also reveal the differences
in both event generators, how they handle the generation processes.
Although both are the general-purpose event generators to simulate
the high-energy lepton-lepton, lepton-hadron and hadron-hadron collisions,
PYTHIA 8 uses Lund String Model for the hadronization process and
has $p_{T}$ ordered parton showers while HERWIG 7 uses a cluster
model to describe the hadronization process based on non-perturbative
gluon splitting and angular-ordered parton showers for the initial-
and final-state QCD jet evolution. HERWIG 7 also has improved leading-logarithmic
(LL) parton showers and colour-cohorence effects \citep{CMS-AzDecPaper}.
In previous experiments it is shown that colour cohorence can effect
the ratio of the difference in pseudo-rapidity over the difference
in azimuthal angle between leading jets \citep{Lee}. Due to the properties
of HERWIG, we see a clearer decorrelation effect with the obtained
observables. Note that MC modelling uncertainty is the dominant systematic
uncertainty in the jet energy detection and both ATLAS and CMS detectors
have recently used above mentioned event generators for forward-backward
jet calibration taking the full difference between PYTHIA 8 and HERWIG
7 as the uncertainty. As a preliminary study to FCC, we have executed
the related simulations in the detector and generator level without
systematic uncertainities. After 36.2 million events in total are
produced and analysed , we were able to obtain clear signals for MN-jet
decorrelations and partially compatible results between two generators.
However for $|\Delta\eta|>4$, one can predict MC modelling differences
in measuring $C_{2}/C_{1}$ and $C_{3}/C_{2}$ at FCC-ep. One can
ignore the entries for pseudo-rapidity interval $|\Delta\eta|>9$
due to low statistics and inconsistency and should also consider the
azimuthal angular jet decorrelations at FCC-hh \citep{Acta}, however
higher decorrelation signs for dijets at FCC-ep is possible due to
the asymmetric collisions on which the experimental data is expected
to be earlier \citep{FCC-dataCol}.}

\begin{table}
\begin{centering}
\caption{Relative uncertainities after separate runs of PDF sets for each $\Delta\eta$
regions}
\par\end{centering}
\begin{centering}
\begin{tabular}{|c|c|c|c|c|c|}
\hline 
\% & PDF Variation & Normalization Scale & $\Delta P_{T}^{mean}$ & $\Delta\eta^{mean}$ & Jet Energy Scale\tabularnewline
\hline 
\hline 
\textcolor{black}{$0>|\Delta\eta|>3$} & \multirow{3}{*}{+ 3.02 - 3.65} & $\pm$0.28 & $\pm$0.72 & $\pm$0.78 & $\pm$0.60\tabularnewline
\cline{1-1} \cline{3-6} \cline{4-6} \cline{5-6} \cline{6-6} 
\textcolor{black}{$3>|\Delta\eta|>6$} &  & $\pm$0.26 & $\pm$0.84 & $\pm$0.25 & $\pm$0.80\tabularnewline
\cline{1-1} \cline{3-6} \cline{4-6} \cline{5-6} \cline{6-6} 
\textcolor{black}{$6>|\Delta\eta|>9$} &  & < $\pm$0.01 & $\pm$0.05 & $\pm$0.01 & $\pm$0.04\tabularnewline
\hline 
\end{tabular}
\par\end{centering}
\end{table}


\begin{thebibliography}{99}
\bibitem{atlas-new-1}ATLAS Collaboration, \textquotedblleft Measurements
of jet vetoes and azimuthal decorrelations in dijet events produced
in pp collisions at sqrt(s)=7 TeV using the ATLAS detector.\textquotedblright{}
The European physical journal. C, Particles and fields vol. 74,11
(2014): 3117. doi:10.1140/epjc/s10052-014-3117-7

\bibitem{atlas-new}ATLAS Collaboration, \textquotedblleft Measurement
of dijet production with a veto on additional central jet activity
in pp collisions at sqrt(s)=7 TeV using the ATLAS detector\textquotedblright ,
JHEP 09 (2011) 053, doi:10.1007/JHEP09(2011)053, arXiv:1107.1641.

\bibitem{CMS-kFactor}CMS Collaboration, \textquotedblleft Ratios
of dijet production cross sections as a function of the absolute difference
in rapidity between jets in proton-proton collisions at sqrt(s) =
7 TeV\textquotedblright , Eur. Phys. J. C 72 (2012) 2216, doi:10.1140/epjc/s10052-012-2216-6,
arXiv:1204.0696.

\bibitem{CMS-AzDecPaper}The CMS collaboration, \textquotedblleft Azimuthal
decorrelation of jets widely separated in rapidity in pp collisions
at \textsurd s= 7 TeV\textquotedblright , J. High Energ. Phys. (2016)
2016: 139. https://doi.org/10.1007/JHEP08(2016)139.

\bibitem{D0-1}D0 Collaboration, \textquotedblleft The azimuthal decorrelation
of jets widely separated in rapidity\textquotedblright , Phys. Rev.
Lett. 77 (1996) 595, doi:10.1103/PhysRevLett.77.595, arXiv:hep-ex/9603010.

\bibitem{D0-2}D0 Collaboration, \textquotedblleft Probing BFKL Dynamics
in the Dijet Cross Section at Large Rapidity Intervals in ppbar Collisions
at sqrt\{s\}=1800 and 630 GeV\textquotedblright , Phys. Rev. Lett.
84 (2000) 5722, doi: 10.1103/PhysRevLett.84.5722, arXiv:hep-ex/9912032. 

\bibitem{FCC-2}\textcolor{black}{A. Ball et al. , Future Circular
Collider Study Hadron Collider Parameters, FCC-ACC-SPC- 0001 (2014).}

\bibitem{CEPC}\textcolor{black}{The CEPC-SPPC Study Group, CEPC-SPPC
Preliminary Conceptual Design Report, Volume II-Accelerator, IHEP-AC-2015-01,
March 2015.}

\bibitem{Acta}\textcolor{black}{I. Ho\c{s}, H. Sayg\i n, S. Kuday,
Azimuthal Angular Decorrelation of Jets at Future High-energy Colliders,
Acta Phys. Pol. B 50, 149 (2019).}

\bibitem{herwig}Bellm, Johannes et al., Herwig 7.0/Herwig++ 3.0 release
note, Eur.Phys.J. C76 (2016) no.4, 196, DOI: 10.1140/epjc/s10052-016-4018-8.

\bibitem{pythia8}T. Sjostrand et al, \textquotedblleft An Introduction
to PYTHIA 8.2\textquotedblright , Comput. Phys. Commun. 191 (2015)
159, DOI: 10.1016/j.cpc.2015.01.024.

\bibitem{FCC-ep}\textcolor{black}{Future Circular Collider Study.
Volume 3: The Hadron Collider (FCC-hh) Conceptual Design Report, preprint
edited by M. Benedikt et al. CERN accelerator reports, CERN-ACC-2018-0058,
Geneva, December 2018. Published in Eur. Phys. J. ST.}

\bibitem{NNpdf23}R. D. Ball, V. Bertone, S. Carrazza, L. Del Debbio,
S. Forte, A. Guffanti, N. P. Hartland, J. Rojo. Parton distributions
with QED corrections, Nuclear Physics B, Volume 877, Issue 2, Pages
290-320, 2013.

\bibitem{NNpdf40}Ball, R.D., Carrazza, S., Cruz-Martinez, J. et al.
The path to proton structure at 1\% accuracy. Eur. Phys. J. C 82,
428 (2022).

\bibitem{FCC-CDR}Abada, A., Abbrescia, M., AbdusSalam, S.S. et al.
FCC Physics Opportunities. Eur. Phys. J. C 79, 474 (2019). https://doi.org/10.1140/epjc/s10\textcolor{black}{052-019-6904-3.}

\bibitem{FCC-1}\textcolor{black}{The Future Circular Collider Study
Group, Kickoff Meeting, 12-15 February 2014, University of Geneva,
Switzerland, https://indico.cern.ch/event/2 82344/. More information
is available on the FCC Web site: http://cern.ch/fcc.}

\bibitem{BFKL} E. A. Kuraev, L. N. Lipatov and V. S. Fadin, \textquotedblleft Multi-reggeon
processes in the Yang-Mills theory\textquotedblright , Sov. Phys.
JETP 44 (1976) 443.

\bibitem{BFKL-2} E. A. Kuraev, L. N. Lipatov and V. S. Fadin, \textquotedblleft The
Pomeranchuk singularity in nonabelian gauge theories\textquotedblright ,
Sov. Phys. JETP 45 (1977) 199.

\bibitem{BFKL-3}I. I. Balitsky and L. N. Lipatov, \textquotedblleft The
Pomeranchuk singularity in quantum chromodynamics\textquotedblright ,
Sov. J. Nucl. Phys. 28 (1978) 822.

\bibitem{DGLAP-1}\textcolor{black}{Altarelli, G. and G. Parisi, ``Asymptotic
Freedom in Parton Language'', Nucl. Phys., B126:298\textendash 318,
1977.}

\bibitem{DGLAP-2}Dokshitzer Y. L., Sov. Phys. JETP, 46(1977) 641,
{[}Zh. Eksp. Teor. Fiz.73, 1216 (1977){]}.

\bibitem{DGLAP-3}M. M. Block, L. Durand, P. Ha, and D.W. McKay, ``Analytic
solution to leading order coupled DGLAP evolution equations: A new
perturbative QCD tool'' Phys. Rev. D 83, 054009 (2011).

\bibitem{Ducloue-1}B. Ducloue, L. Szymanowski, and S. Wallon, Confronting
Mueller-Navelet jets in NLL BFKL with LHC experiments at 7 TeV, J.
High Energy Phys. 05 (2013) 096

\bibitem{Ducloue-2}B. Ducloue, L. Szymanowski, and S. Wallon, Evidence
for High-Energy Resummation Effects in Mueller-Navelet Jets at the
LHC, Phys. Rev. Lett. 112, 082003 (2014)

\bibitem{D0-7}D0 Collaboration, B. Abbottet al,Phys. Rev. Lett.84(2000)
5722

\bibitem{Duca}\textcolor{black}{V. del Duca and C. R. Schmidt, \textquotedblleft Mini
- jet corrections to Higgs production,\textquotedblright{} Phys. Rev.
D: Part. Fields 49, 177 (1994); hep-ph/9305346. https://doi.org/10.1103/PhysRevD.49.177}

\bibitem{Stirling}\textcolor{black}{W. J. Stirling, \textquotedblleft Production
of jet pairs at large relative rapidity in hadron hadron collisions
as a probe of the perturbative pomeron,\textquotedblright{} Nucl.
Phys. B 423, 56 (1994); hep-ph/9401266. https://doi.org/10.1016/0550-3213(94)90565-7}

\bibitem{murzin}\textcolor{black}{Murzin, V.A. Dijets with Large
Rapidity Separation at CMS. Phys. Part. Nuclei Lett. 16, 469\textendash 474
(2019). https://doi.org/10.1134/S1547477119050224 }

\bibitem{madgraph}J. Alwall et al, \textquotedbl The automated computation
of tree-level and next-to-leading order differential cross sections,
and their matching to parton shower simulations\textquotedbl , arXiv:1405.0301
{[}hep-ph{]}

\bibitem{anti-kT}M. Cacciari, G. P. Salam, and G. Soyez, \textquotedblleft The
anti-kt jet clustering algorithm\textquotedblright , JHEP 04 (2008)
063, doi:10.1088/1126-6708/2008/04/063, arXiv:0802.1189.

\bibitem{fastjet}Cacciari, M., Salam, G.P. \& Soyez, \textquotedblleft FastJet
User Manual\textquotedblright , G. Eur. Phys. J. C (2012) 72: 1896.
https://doi.org/10.1140/epjc/s10052-012-1896-2.

\bibitem{Delphes}The DELPHES 3 collaboration., de Favereau, J., Delaere,
C. et al. DELPHES 3: a modular framework for fast simulation of a
generic collider experiment. J. High Energ. Phys. 2014, 57 (2014).
https://doi.org/10.1007/JHEP02(2014)057

\bibitem{NewDelphes}A. Mertens, New features in Delphes 3. J. Phys.:
Conf. Ser. 608 012045, 2015.

\bibitem{Lee}Lee, Jason. Multi-jet correlations and colour coherence
phenomena. EPJ Web o\textcolor{black}{f Conferences. 141. 04003. 10.1051/epjconf/201714104003.}

\bibitem{FCC-dataCol}\textcolor{black}{M. Klein, ``The Development
of the LHeC/PERLE/FCC-eh Project'', The LHeC/FCCeh and PERLE Workshop
at Orsay, October 2022, https://indico.ijclab.in2p3.fr/event/8623/contributions/27078/attachments/19800/27213/projectMK.pdf}
\end{thebibliography}
\end{document}